\documentclass[12pt,a4paper]{article}

\usepackage{amsfonts}
\usepackage{graphicx}
\usepackage{amsmath}
\usepackage{amsthm}
\usepackage{longtable}
\usepackage{float}

\theoremstyle{definition}
\newtheorem{definition}{Definition}
\newtheorem{theorem}{Theorem}

\begin{document}

\title{Higher order singular value decomposition and the reduced density matrices of three qubits}

\author{Choong Pak Shen$^1$ \and Hishamuddin Zainuddin$^{1,2,3}$ \and Chan Kar Tim$^1$ \and Sh. K. Said Husain$^1$}
\date{$^1$ Institute for Mathematical Research, Universiti Putra Malaysia, 43400 Serdang, Selangor, Malaysia.\\
$^2$ Malaysia-Italy Center for Mathematical Sciences (MICEMS), Universiti Putra Malaysia, 43400 Serdang, Selangor, Malaysia.\\
$^3$ Corresponding author: \texttt{hisham@upm.edu.my}\\
Date: \today}

\maketitle

\begin{abstract}
In this paper, we demonstrate that higher order singular value decomposition (HOSVD) can be used to identify special states in three qubits by local unitary (LU) operations. Since the matrix unfoldings of three qubits are related to their reduced density matrices, HOSVD simultaneously diagonalizes the one-body reduced density matrices of three qubits. From the all-orthogonality conditions of HOSVD, we computed the special states of three qubits. Furthermore, we showed that it is possible to construct a polytope that encapsulates all the special states of three qubits by LU operations with HOSVD.
\end{abstract}

\section{Introduction}
Being the central characteristic of composite quantum systems, entanglement has been studied extensively in the past from various perspectives, such as the classification of multipartite states \cite{Acin2000,Carteret2000,Albeverio2005,Liu2012,Li2013,Li2014}, the geometry of quantum state space \cite{Kus2001,Sinolecka2002,Bernevig2003} and more recently, the resource-theoretic \cite{Bennett1998,Eltschka2014,Coecke2016,Chitambar2018} and categorical approach \cite{Coecke2004,Coecke2010}. Apart from the fact that entanglement connects deeply to the foundations of quantum theory, it can be utilized as a resource in quantum information processing. From this perspective, it is important to be able to quantify entanglement and classify entangled states based on the computational tasks they can perform.

The Hilbert space of a composite quantum system is described by the tensor product of its subsystems' Hilbert spaces. This tensor product structure naturally endows tensorial properties to the elements of multipartite states, thus allowing us to employ multilinear algebraic methods on them. As an example, we can apply singular value decomposition (SVD) on the elements of bipartite states and restate it as Schmidt decomposition \cite{Nielsen2000}, which is a widely used approach in the local unitary (LU) classification of bipartite states. It is also known that the Schmidt coefficients are LU invariants of the entanglement classes for bipartite states \cite{Carteret2000,Sinolecka2002}.

Given the successful precedence in bipartite states, it is natural to consider Schmidt decomposition in the LU classification of multipartite states. This idea turned out to be unsuccessful \cite{Peres1995} since multipartite states cannot be generally represented by only the Schmidt coefficients. If one were to follow a similar concept, Schmidt decomposition or equivalently SVD has to be generalized.

As a matter of fact, such a generalization has been considered in the mathematical literature back in 2000 \cite{Lathauwer2000}. Particularly, the requirement of matrix diagonalization in SVD is relaxed. This generalized version of SVD is called higher order singular value decomposition (HOSVD). It is applicable to higher order tensors, which is the tensorial representation of multipartite states. The idea of utilizing HOSVD in the LU classification of multipartite pure states was first suggested in \cite{Liu2012}. Subsequently, a general scheme was proposed in \cite{Li2013} to identify the LU equivalence between two multipartite pure states. Later, Li et. al. \cite{Li2014} presented a necessary and sufficient criterion to check if two multipartite mixed states are local unitary equivalent or not.

In this paper, we choose a different approach of utilizing HOSVD in finding the special states of three qubits. Instead of focusing on the local symmetries \cite{Liu2012,Li2013,Li2014} of multipartite states, we make use of the properties of matrix unfolding and HOSVD to identify the special states of three-qubit pure states by LU operations. We begin by defining the multilinear algebraic equivalence of three-qubit states in Section \ref{SecTensors}. Then, we discuss the matrix unfolding of tensors in Section \ref{SecMatrixUnfolding}. Here, we found that the matrix unfoldings of three-qubit tensors are related to their reduced density matrices. The definition of HOSVD is presented in Section \ref{SecHOSVD}, whereby HOSVD simply defines and guarantees the existence of simultaneous diagonalization of one-body reduced density matrices for three qubits. Based on the all-orthogonality conditions of HOSVD, we calculated the special states of three qubits and constructed a polytope of three qubits by LU operations in Section \ref{SecResults}. The special states of three qubits are in correspondence to an earlier work by \cite{Carteret2000}.

\section{Tensors} \label{SecTensors}

Tensors are indexed mathematical objects coming from the tensor product of vector spaces and can be regarded as multi-dimensional arrays \cite{Kolda2009}. Let $\mathcal{X} = [\chi_{i_1 i_2 \ldots i_n \ldots i_N}] \in V^{I_1} \otimes V^{I_2} \otimes \ldots \otimes V^{I_n} \otimes \ldots \otimes V^{I_N}$ be a tensor, where $V^{I_n}$ is the $n$-th vector space of dimension $I_n$. The total number of indices $N$ of a tensor $\mathcal{X}$ is called the order of a tensor. Thus, tensors of order 1 are vectors while tensors of order 2 are matrices. For tensors of order 3 and above, we call them higher order tensors. The Hilbert space of three qubits, for instance, is the tensor product of three complex vector spaces, $H = \mathbb{C}^2 \otimes \mathbb{C}^2 \otimes \mathbb{C}^2$. Therefore, any three-qubit states, or tripartite states in general, are tensors of order 3.

Now, let $\mathcal{V}_n = [\nu_{i_n}] \in V^{I_n}$ be vectors (or tensors of order 1) in the $n$-th vector space $V^{I_n}$ of dimension $I_n$. An $N$-th order tensor $\mathcal{X} = [\chi_{i_1 i_2 \ldots i_n \ldots i_N}] \in V^{I_1} \otimes V^{I_2} \otimes \ldots \otimes V^{I_n} \otimes \ldots \otimes V^{I_N}$ is of rank 1 if it can be written as
\begin{equation}
\mathcal{X} = \mathcal{V}_1 \otimes \mathcal{V}_2 \otimes \ldots \otimes \mathcal{V}_n \otimes \ldots \otimes \mathcal{V}_N,
\end{equation}
or equivalently if its element $\chi_{i_1 i_2 \ldots i_n \ldots i_N}$ can be written as
\begin{equation}
\chi_{i_1 i_2 \ldots i_n \ldots i_N} = \nu_{i_1} \nu_{i_2} \ldots \nu_{i_n} \ldots \nu_{i_N}.
\end{equation}
Rank 1 tensors are also called simple tensors \cite{Sawicki2013}. Now, let $\mathcal{X}_r$ be a $N$-th order tensor of rank 1. Then, the rank of a generic $N$-th order tensor $\mathcal{X}$ is the minimum number $R$ of rank 1 tensors $\mathcal{X}_r$ combined linearly to form $\mathcal{X}$ \cite{Lim2013},
\begin{equation}
\text{Rank}(\mathcal{X}) := \text{Min} \left\{ R: \mathcal{X} = \sum^R_{r=1} \mathcal{X}_r \right\}.
\end{equation}

As an example, the GHZ state
\begin{equation*}
\left| \text{GHZ} \right\rangle = \psi_{111} \left| 111 \right\rangle + \psi_{222} \left| 222 \right\rangle
\end{equation*}
is a third order tensor of rank 2, while the W state
\begin{equation*}
\left| \text{W} \right\rangle = \psi_{112} \left| 112 \right\rangle + \psi_{121} \left| 121 \right\rangle + \psi_{211} \left| 211 \right\rangle
\end{equation*}
is a third order tensor of rank 3 \cite{Sawicki2013}. Even though the idea of tensor rank is not the main focus of this paper, we would like to highlight that tensor rank is related to the transformation of tripartite entangled states through stochastic local operation and classical communication (SLOCC) \cite{Chitambar2008} and can be an algebraic measure of entanglement \cite{Brylinski2002}.

\section{Matrix unfolding of higher order tensors} \label{SecMatrixUnfolding}
\subsection{Matrix unfolding and local transformation of tensors}

While it is possible to write down a higher order tensor by listing its tensor elements, it will be more convenient to devise a standardized way of representing a higher order tensor as matrices. Such a method is called matrix unfolding \cite{Lathauwer2000}.

\begin{definition}[Matrix unfolding] \label{MatrixUnfoldingDef}

\noindent \emph{Let $\mathcal{X} \in V^{I_1} \otimes V^{I_2} \otimes \ldots \otimes V^{I_n} \otimes \ldots \otimes V^{I_N}$ be an $N$-th order tensor, where $V^{I_n}$ is the $n$-th vector space of dimension $I_n$. The $n$-th matrix unfolding, $X_{(n)}$, is a matrix of size $I_n \times (I_{n+1} I_{n+2} \ldots I_N I_1 I_2 \ldots I_{n-1})$, whereby the tensor element $\chi_{i_1 i_2 \ldots i_n \ldots i_N}$ will be located at the position with row index $i_n$ and column index
\begin{align}
& (i_{n+1} -1) I_{n+2} I_{n+3} \ldots I_N I_1 I_2 \ldots I_{n-1} \nonumber \\ & \quad + (i_{n+2} -1) I_{n+3} I_{n+4} \ldots I_N I_1 I_2 \ldots I_{n-1} + \ldots \nonumber \\ & \quad + (i_N -1) I_1 I_2 \ldots I_{n-1} + (i_1 -1) I_2 I_3 \ldots I_{n-1} \nonumber \\ & \quad + (i_2 -1) I_3 I_4 \ldots I_{n-1} + \ldots + i_{n-1}.
\end{align}}
\end{definition}

When the elements of multipartite states are represented as tensors, the matrix unfolding of higher order tensors allows us to define local transformation of multipartite states \cite{Li2013} in a convenient way, as follow.

\begin{definition}[Local transformation of tensors] \label{LocalTransformationDef}

\noindent \emph{Let $\mathcal{X} \in V^{I_1} \otimes V^{I_2} \otimes \ldots \otimes V^{I_n} \otimes \ldots \otimes V^{I_N}$ be an $N$-th order tensor, where $V^{I_n}$ is the $n$-th vector space of dimension $I_n$. Let $\mathcal{M}^{(n)} \in \text{GL}(V^{I_n})$ be the linear transformation matrix on the vector space $V^{I_n}$. Then, the local transformation of an $N$-th order tensor $\mathcal{X}$ is given as
\begin{equation} \label{LocalTransformation}
\mathcal{X}' = \mathcal{M}^{(1)} \otimes \mathcal{M}^{(2)} \otimes \ldots \otimes \mathcal{M}^{(n)} \otimes \ldots \otimes \mathcal{M}^{(N)} \mathcal{X},
\end{equation}
where $\mathcal{M}^{(1)} \otimes \mathcal{M}^{(2)} \otimes \ldots \otimes \mathcal{M}^{(N)} \in \text{GL}(V^{I_1}) \times \text{GL}(V^{I_2}) \times \ldots \times \text{GL}(V^{I_N})$. The $n$-th matrix unfolding of equation (\ref{LocalTransformation}) can be written as
\begin{align} \label{MatrixTensorMultiplication}
X'_{(n)} = \mathcal{M}^{(n)} X_{(n)} & \left[\mathcal{M}^{(n+1)} \otimes \ldots \otimes \mathcal{M}^{(N)} \otimes \mathcal{M}^{(1)} \otimes \ldots \otimes \mathcal{M}^{(n-1)}\right]^T,
\end{align}
where the superscript $T$ denotes matrix transpose.}
\end{definition}

\subsection{The matrix unfolding of three qubits}

We shall now demonstrate the procedure of matrix unfolding for the case of three qubits. Given the three-qubit state,
\begin{equation}
\left| \psi \right\rangle = \sum_{i_1, i_2, i_3 = 1}^2 \psi_{i_1 i_2 i_3} \left| i_1 i_2 i_3 \right\rangle,
\end{equation}
one can denote its tensorial form as $\Psi = [\psi_{i_1 i_2 i_3}] \in H = \mathbb{C}^2 \otimes \mathbb{C}^2 \otimes \mathbb{C}^2$. From Definition \ref{MatrixUnfoldingDef}, the 1-, 2- and 3-matrix unfoldings of $\Psi$ are given as the following:-
\begin{itemize}
\item{
The first matrix unfolding, $\Psi_{(1)}$, is an $I_1 \times (I_2 I_3) = 2 \times 4$ matrix with tensor elements $\psi_{i_1 i_2 i_3}$ situated at position with row index $i_1$ and column index $(i_2 - 1) I_3 + i_3 = 2 (i_2 - 1) + i_3$,
\begin{equation}
\Psi_{(1)}= 
\begin{pmatrix}
\psi_{111} & \psi_{112} & \psi_{121} & \psi_{122} \\
\psi_{211} & \psi_{212} & \psi_{221} & \psi_{222}
\end{pmatrix}.
\end{equation}
}

\item{
The second matrix unfolding, $\Psi_{(2)}$, is an $I_2 \times (I_3 I_1) = 2 \times 4$ matrix with tensor elements $\psi_{i_1 i_2 i_3}$ situated at position with row index $i_2$ and column index $(i_3 - 1) I_1 + i_1 = 2 (i_3 - 1) + i_1$,
\begin{equation}
\Psi_{(2)}= 
\begin{pmatrix}
\psi_{111} & \psi_{211} & \psi_{112} & \psi_{212} \\
\psi_{121} & \psi_{221} & \psi_{122} & \psi_{222}
\end{pmatrix}.
\end{equation}
}

\item{
The third matrix unfolding, $\Psi_{(3)}$, is an $I_3 \times (I_1 I_2) = 2 \times 4$ matrix with tensor elements $\psi_{i_1 i_2 i_3}$ situated at position with row index $i_3$ and column index $(i_1 - 1) I_2 + i_2 = 2 (i_1 - 1) + i_2$,
\begin{equation}
\Psi_{(3)}= 
\begin{pmatrix}
\psi_{111} & \psi_{121} & \psi_{211} & \psi_{221} \\
\psi_{112} & \psi_{122} & \psi_{212} & \psi_{222}
\end{pmatrix}.
\end{equation}
}
\end{itemize}

These matrix unfoldings have close resemblance to the matrices $\Psi_{A|BC}$, $\Psi_{B|CA}$, and $\Psi_{C|AB}$ used in the literature (see, for example \cite{Lamata2007}). By direct comparison with the reduced density matrices of three qubits, we find that
\begin{align}
\rho^{AB} & = \Psi_{(3)}^T \bar{\Psi}_{(3)}, \\
\rho^{CA} & = \Psi_{(2)}^T \bar{\Psi}_{(2)}, \\
\rho^{BC} & = \Psi_{(1)}^T \bar{\Psi}_{(1)}, \\
\rho^{A} & = \Psi_{(1)} \Psi_{(1)}^\dagger, \label{ThreeQubitsReducedDensityMatrix1} \\
\rho^{B} & = \Psi_{(2)} \Psi_{(2)}^\dagger, \label{ThreeQubitsReducedDensityMatrix2} \\
\rho^{C} & = \Psi_{(3)} \Psi_{(3)}^\dagger, \label{ThreeQubitsReducedDensityMatrix3}
\end{align}
where overhead bar denotes complex conjugate while superscript $\dagger$ denotes conjugate transpose.

For three qubits, its local unitary (LU) transformation can be defined as the action of the LU operators $U^{(1)} \otimes U^{(2)} \otimes U^{(3)} \in \text{SU}(2) \times \text{SU}(2) \times \text{SU}(2)$ acting on the three-qubit state,
\begin{align} \label{LocalUnitaryTransformationThreeQubits}
U^{(1)} \otimes U^{(2)} \otimes U^{(3)} \left| \psi \right\rangle = \sum_{j_1, j_2, j_3 = 1}^2 \sum_{i_1, i_2, i_3 = 1}^2 u^{(1)}_{j_1 i_1} u^{(2)}_{j_2 i_2} u^{(3)}_{j_3 i_3} \psi_{i_1 i_2 i_3} \left| j_1 j_2 j_3 \right\rangle.
\end{align}
From Definition \ref{LocalTransformationDef}, the tensorial form of equation (\ref{LocalUnitaryTransformationThreeQubits}) can be rewritten as
\begin{equation} \label{LocalUnitaryTransformationThreeQubitsTensor}
\Psi' = U^{(1)} \otimes U^{(2)} \otimes U^{(3)} \Psi.
\end{equation}
From equations (\ref{ThreeQubitsReducedDensityMatrix1}) and (\ref{LocalUnitaryTransformationThreeQubitsTensor}), we can show that
\begin{align}
{\rho'}^{A} & = \Psi'_{(1)} {\Psi'}_{(1)}^{\dagger} \nonumber \\ & = U^{(1)} \Psi_{(1)} \left[U^{(2)} \otimes U^{(3)}\right]^T \left\{U^{(1)} \Psi_{(1)} \left[U^{(2)} \otimes U^{(3)}\right]^T\right\}^\dagger \nonumber \\ & = U^{(1)} \Psi_{(1)} \left[U^{(2)} \otimes U^{(3)}\right]^T \overline{\left[U^{(2)} \otimes U^{(3)}\right]} \Psi_{(1)}^\dagger U^{(1)\dagger} \nonumber \\ & = U^{(1)} \Psi_{(1)} \Psi_{(1)}^\dagger U^{(1)\dagger} \nonumber \\ & = U^{(1)} \rho^A U^{(1)\dagger}. \label{LUOnRDM1}
\end{align}
Similar procedure can be performed on equations (\ref{ThreeQubitsReducedDensityMatrix2}) and (\ref{ThreeQubitsReducedDensityMatrix3}) to get equations (\ref{LUOnRDM2}) and (\ref{LUOnRDM3}), respectively:-
\begin{align}
{\rho'}^{B} & = U^{(2)} \rho^B U^{(2)\dagger}, \label{LUOnRDM2} \\
{\rho'}^{C} & = U^{(3)} \rho^C U^{(3)\dagger}. \label{LUOnRDM3}
\end{align}
The simple exercise above shows that equation (\ref{LocalUnitaryTransformationThreeQubitsTensor}) is indeed a LU action and the one-body reduced density matrices ${\rho'}^{A}, {\rho'}^{B}, {\rho'}^{C}$will fall under the same LU equivalence classes as $\rho^{A}, \rho^{B}, \rho^{C}$.

\section{Higher order singular value decomposition} \label{SecHOSVD}
\subsection{Definition}

Having defined tensors and matrix unfolding, our next goal is to introduce a type of tensor decomposition called higher order singular value decomposition (HOSVD) \cite{Lathauwer2000,Li2013}, which is the generalized version of singular value decomposition (SVD).

\begin{theorem}[Higher order singular value decomposition] \label{HOSVDTheo}

\noindent \emph{Let $\mathcal{X} \in \mathbb{C}^{I_1} \otimes \mathbb{C}^{I_2} \otimes \ldots \otimes \mathbb{C}^{I_n} \otimes \ldots \otimes \mathbb{C}^{I_N}$ be an $N$-th order complex tensor, where $\mathbb{C}^{I_n}$ is the $n$-th complex vector space of dimension $I_n$. There exists a core tensor $\mathcal{T}$ of $\mathcal{X}$ and a set of unitary matrices $U^{(1)},\, U^{(2)},\, \ldots,\, U^{(n)},\, \ldots,\, U^{(N)}$ such that
\begin{equation} \label{HOSVDEquation}
\mathcal{X} = U^{(1)} \otimes U^{(2)} \otimes \ldots \otimes U^{(n)} \otimes \ldots \otimes U^{(N)} \mathcal{T}.
\end{equation}
The core tensor $\mathcal{T}$ is also an $N$-th order complex tensor for which its subtensors $\mathcal{T}_{i_{n}=\alpha}$, obtained by fixing the $n$-th index to $\alpha$, have the properties of
\begin{enumerate}
\item{
All-orthogonality: Two subtensors $\mathcal{T}_{i_{n}=\alpha}$ and $\mathcal{T}_{i_{n}=\beta}$ are orthogonal for all possible values of $n$, $\alpha$ and $\beta$, subject to $\alpha \neq \beta$:
\begin{align} \label{AllOrthogonality}
\left \langle \mathcal{T}_{i_{n}=\alpha}, \mathcal{T}_{i_{n}=\beta} \right \rangle & = \sum_{i_1 i_2 \ldots i_{n-1} i_{n+1} \ldots i_N} \bar{t}_{i_1 i_2 \ldots i_{n-1} \alpha i_{n+1} \ldots i_N} t_{i_1 i_2 \ldots i_{n-1} \beta i_{n+1} \ldots i_N} \nonumber \\ & = 0 \; \; \text{when} \; \; \alpha \neq \beta;
\end{align}
}
\item{
Ordering:
\begin{equation}
\left| \mathcal{T}_{i_{n}=1} \right| \geq \left| \mathcal{T}_{i_{n}=2} \right| \geq \ldots \geq \left| \mathcal{T}_{i_{n}=I_n} \right| \geq 0
\end{equation}
for all possible values of $n$,
}
\end{enumerate}
where $t_{i_1 i_2 \ldots i_N}$ is the element of the tensor $\mathcal{T}$ and $\left| \mathcal{T}_{i_n=i} \right| = \sqrt{\left\langle \mathcal{T}_{i_{n}=i}, \mathcal{T}_{i_{n}=i} \right\rangle}$ is called the $n$-mode singular value of $\mathcal{X}$, $\sigma_i^{(n)}$.}
\end{theorem}

Due to Definition \ref{LocalTransformationDef}, equation (\ref{HOSVDEquation}) can be rewritten as
\begin{align}
X_{(n)} = U^{(n)} T_{(n)} & \left[U^{(n+1)} \otimes U^{(n+2)} \otimes \ldots \otimes U^{(N)} \right. \nonumber \\ & \quad \left. \otimes U^{(1)} \otimes U^{(2)} \otimes \ldots \otimes U^{(n-1)}\right]^T,
\end{align}
where $X_{(n)}$ and $T_{(n)}$ are $I_n \times (I_{n+1} I_{n+2} \ldots I_N I_1 I_2 \ldots I_{n-1})$-complex matrices, and $U^{(n)}$ are unitary matrices of size $I_n \times I_n$.

As stated in \cite{Lathauwer2000}, SVD reduces any real or complex matrix into a diagonal matrix $\Lambda$ of real entries, whereas HOSVD relaxes this property into a set of all-orthogonality conditions (\ref{AllOrthogonality}). To put this into perspective, instead of requiring the matrix $\Lambda$ to be diagonal, HOSVD only requires that the row and column vectors of $\Lambda$ to be orthogonal to each other. In this sense, HOSVD generalizes SVD.

\subsection{Higher order singular value decomposition on three qubits} \label{SubSecHOSVDThreeQubits}

From equation (\ref{AllOrthogonality}), the core tensor elements of three qubits $\mathcal{T}_\psi = \left[ t_{i_1 i_2 i_3} \right]$ satisfy the following all-orthogonality conditions,
\begin{align}
\bar{t}_{111} t_{211} + \bar{t}_{121} t_{221} + \bar{t}_{112} t_{212} + \bar{t}_{122} t_{222} & = 0, \label{ThreeQubitsAllOrthogonality1} \\
\bar{t}_{111} t_{121} + \bar{t}_{211} t_{221} + \bar{t}_{112} t_{122} + \bar{t}_{212} t_{222} & = 0, \label{ThreeQubitsAllOrthogonality2} \\
\bar{t}_{111} t_{112} + \bar{t}_{211} t_{212} + \bar{t}_{121} t_{122} + \bar{t}_{221} t_{222} & = 0. \label{ThreeQubitsAllOrthogonality3}
\end{align}

The $n$-mode singular values are given as
\begin{align}
\sigma_1^{(1)} & = \sqrt{\left| t_{111} \right|^2 + \left| t_{112} \right|^2 + \left| t_{121} \right|^2 + \left| t_{122} \right|^2}, \label{ThreeQubitsFirstModeSingularValues} \\
\sigma_2^{(1)} & = \sqrt{\left| t_{211} \right|^2 + \left| t_{212} \right|^2 + \left| t_{221} \right|^2 + \left| t_{222} \right|^2}, \\
\sigma_1^{(2)} & = \sqrt{\left| t_{111} \right|^2 + \left| t_{112} \right|^2 + \left| t_{211} \right|^2 + \left| t_{212} \right|^2}, \label{ThreeQubitsSecondModeSingularValues} \\
\sigma_2^{(2)} & = \sqrt{\left| t_{121} \right|^2 + \left| t_{122} \right|^2 + \left| t_{221} \right|^2 + \left| t_{222} \right|^2}, \\
\sigma_1^{(3)} & = \sqrt{\left| t_{111} \right|^2 + \left| t_{121} \right|^2 + \left| t_{211} \right|^2 + \left| t_{221} \right|^2}, \label{ThreeQubitsThirdModeSingularValues} \\
\sigma_2^{(3)} & = \sqrt{\left| t_{112} \right|^2 + \left| t_{122} \right|^2 + \left| t_{212} \right|^2 + \left| t_{222} \right|^2}.
\end{align}

The normalization condition of probability amplitudes tells us that the square of the singular values for a particular matrix unfolding should sum up to be 1, i.e.
\begin{align}
\sigma_1^{(1)2} + \sigma_2^{(1)2} & = 1, \\
\sigma_1^{(2)2} + \sigma_2^{(2)2} & = 1, \\
\sigma_1^{(3)2} + \sigma_2^{(3)2} & = 1.
\end{align}
Due to the ordering property of HOSVD, $\sigma_1^{(n)2} \geq \sigma_2^{(n)2}$ for $n = 1, 2, 3$.

By comparison, it is not difficult to see that the all-orthogonality conditions (\ref{ThreeQubitsAllOrthogonality1}), (\ref{ThreeQubitsAllOrthogonality2}) and (\ref{ThreeQubitsAllOrthogonality3}) are the off-diagonal terms of the one-body reduced density matrices $\rho^A$, $\rho^B$ and $\rho^C$ respectively. This means that HOSVD simultaneously diagonalizes the one-body reduced density matrices of three qubits,
\begin{align}
\rho^A & = U^{(1)} \rho_d^A U^{(1)\dagger}, \label{ThreeQubitsSpectralTheorem1} \\
\rho^B & = U^{(2)} \rho_d^B U^{(2)\dagger}, \label{ThreeQubitsSpectralTheorem2} \\
\rho^C & = U^{(3)} \rho_d^C U^{(3)\dagger}, \label{ThreeQubitsSpectralTheorem3}
\end{align}
where
\begin{align}
\rho_d^A = T_{(1)} T_{(1)}^\dagger = \begin{pmatrix}
\sigma_1^{(1)2} & 0 \\
0 & \sigma_2^{(1)2}
\end{pmatrix}, \label{ThreeQubitsHOSVDDiagonalization1} \\
\rho_d^B = T_{(2)} T_{(2)}^\dagger = \begin{pmatrix}
\sigma_1^{(2)2} & 0 \\
0 & \sigma_2^{(2)2}
\end{pmatrix}, \label{ThreeQubitsHOSVDDiagonalization2} \\
\rho_d^C = T_{(3)} T_{(3)}^\dagger = \begin{pmatrix}
\sigma_1^{(3)2} & 0 \\
0 & \sigma_2^{(3)2}
\end{pmatrix}, \label{ThreeQubitsHOSVDDiagonalization3}
\end{align}
and $T_{(1)}$, $T_{(2)}$ and $T_{(3)}$ are the 1-, 2- and 3-matrix unfolding of $\mathcal{T}_\psi$. In this case, we can say that equations (\ref{ThreeQubitsSpectralTheorem1}) to (\ref{ThreeQubitsSpectralTheorem3}) are the spectral theorem of Hermitian matrices \cite{Axler2015} in disguise.  By Definition \ref{LocalTransformationDef}, HOSVD is a local unitary (LU) transformation, hence the three-qubit tensor $\Psi$ and the core tensor $\mathcal{T}_\psi$ are LU equivalent.

\section{Determining special three-qubit states} \label{SecResults}

\subsection{All-orthogonality conditions of three qubits}

Besides diagonalizing the one-body reduced density matrices of three qubits, we found that it is possible to make use of the all-orthogonality conditions to determine the special states of three qubits. To show this, we must first combine the three equations (\ref{ThreeQubitsAllOrthogonality1}), (\ref{ThreeQubitsAllOrthogonality2}) and (\ref{ThreeQubitsAllOrthogonality3}) together. By rearranging equations (\ref{ThreeQubitsAllOrthogonality1}) and (\ref{ThreeQubitsAllOrthogonality2}), we get
\begin{align}
& t_{111} = - \frac{\bar{t}_{221} (t_{121} t_{212} - t_{122} t_{211}) + t_{112} (\left| t_{212} \right|^2 - \left| t_{122} \right|^2)}{t_{212} \bar{t}_{211} - t_{122} \bar{t}_{121}}, \label{t111} \\
& t_{222} = \frac{\bar{t}_{112} (t_{121} t_{212} - t_{122} t_{211}) + t_{221} (\left| t_{121} \right|^2 - \left| t_{211} \right|^2)}{\bar{t}_{212} t_{211} - \bar{t}_{122} t_{121}}. \label{t222}
\end{align}
Substituting the above equations in (\ref{ThreeQubitsAllOrthogonality3}) and comparing the real and imaginary parts, we get
\begin{align}
& \left| t_{112} \right|^2 (\left| t_{122} \right|^2 - \left| t_{212} \right|^2) + \left| t_{121} \right|^2 (\left| t_{221} \right|^2 - \left| t_{122} \right|^2) \nonumber \\ & \qquad + \left| t_{211} \right|^2 (\left| t_{212} \right|^2 - \left| t_{221} \right|^2) = 0, \label{ThreeQubitsSub1} \\
& \bar{t}_{112} \bar{t}_{221} (t_{122} t_{211} - t_{121} t_{212}) + \bar{t}_{121} \bar{t}_{212} (t_{112} t_{221} - t_{122} t_{211}) \nonumber \\ & \qquad + \bar{t}_{122} \bar{t}_{211} (t_{121} t_{212} - t_{112} t_{221}) = 0. \label{ThreeQubitsSub2}
\end{align}
By adding some self-canceling terms, equation (\ref{ThreeQubitsSub1}) becomes
\begin{subequations}
\begin{align}
\left| t_{112} \right|^2 \left[ \sigma_1^{(1)2} - \sigma_1^{(2)2} \right] + \left| t_{211} \right|^2 \left[ \sigma_1^{(2)2} - \sigma_1^{(3)2} \right] + \left| t_{121} \right|^2 \left[ \sigma_1^{(3)2} - \sigma_1^{(1)2} \right] = 0, \label{ThreeQubitsFinalAllOrtho1a}
\end{align}
\begin{align}
\left| t_{221} \right|^2 \left[ \sigma_1^{(1)2} - \sigma_1^{(2)2} \right] + \left| t_{122} \right|^2 \left[ \sigma_1^{(2)2} - \sigma_1^{(3)2} \right] + \left| t_{212} \right|^2 \left[ \sigma_1^{(3)2} - \sigma_1^{(1)2} \right] = 0. \label{ThreeQubitsFinalAllOrtho1b}
\end{align}
\end{subequations}
We note that equations (\ref{ThreeQubitsFinalAllOrtho1a}) and (\ref{ThreeQubitsFinalAllOrtho1b}) are equivalent.

Meanwhile, it is possible to rewrite equation (\ref{ThreeQubitsSub2}) as
\begin{subequations}
\begin{align}
& (t_{112} t_{221} - t_{121} t_{212}) (\bar{t}_{121} \bar{t}_{212} - \bar{t}_{122} \bar{t}_{211}) - \nonumber \\ & \qquad (\bar{t}_{112} \bar{t}_{221} - \bar{t}_{121} \bar{t}_{212}) (t_{121} t_{212} - t_{122} t_{211}) = 0, \label{ThreeQubitsFinalAllOrtho2a}
\end{align}
\begin{align}
& (t_{112} t_{221} - t_{122} t_{211}) (\bar{t}_{121} \bar{t}_{212} - \bar{t}_{112} \bar{t}_{221}) - \nonumber \\ & \qquad (\bar{t}_{112} \bar{t}_{221} - \bar{t}_{122} \bar{t}_{211}) (t_{121} t_{212} - t_{112} t_{221}) = 0, \label{ThreeQubitsFinalAllOrtho2b}
\end{align}
\begin{align}
& (t_{122} t_{211} - t_{121} t_{212}) (\bar{t}_{112} \bar{t}_{221} - \bar{t}_{122} \bar{t}_{211}) - \nonumber \\ & \qquad (\bar{t}_{122} \bar{t}_{211} - \bar{t}_{121} \bar{t}_{212}) (t_{112} t_{221} - t_{122} t_{211}) = 0. \label{ThreeQubitsFinalAllOrtho2c}
\end{align}
\end{subequations}
Similarly, equations (\ref{ThreeQubitsFinalAllOrtho2a}), (\ref{ThreeQubitsFinalAllOrtho2b}) and (\ref{ThreeQubitsFinalAllOrtho2c}) are equivalent.

In Section \ref{SubSecSeparability}, we will discuss the relationship between equations (\ref{ThreeQubitsFinalAllOrtho2a}), (\ref{ThreeQubitsFinalAllOrtho2b}), (\ref{ThreeQubitsFinalAllOrtho2c}) and the separability of three qubits. Meanwhile, the significance of equations (\ref{ThreeQubitsFinalAllOrtho1a}) or (\ref{ThreeQubitsFinalAllOrtho1b}) will be studied in Section \ref{SubSecSpecialStatesThreeQubits}.

\subsection{Separability conditions of three qubits} \label{SubSecSeparability}

The separability of three qubits can be checked by the separability conditions. In order to derive the separability conditions, for example the bi-separability conditions of $C|AB$, we can first define the following states
\begin{align*}
\left| \psi_C \right\rangle & = c_1 \left| 1 \right\rangle + c_2 \left| 2 \right\rangle, \\
\left| \psi_{AB} \right\rangle & = a_{11} \left| 11 \right\rangle + a_{12} \left| 12 \right\rangle + a_{21} \left| 21 \right\rangle + a_{22} \left| 22 \right\rangle.
\end{align*}
With the tensor product of $\left| \psi_{AB} \right\rangle$ and $\left| \psi_C \right\rangle$, one can compare the coefficients and conclude that a three-qubit state is bi-separable with respect to $AB$ and $C$ if it satisfies the following conditions:-
\begin{align}
\psi_{111} \psi_{222} = \psi_{112} \psi_{221}, \label{BiseparableCAB1} \\
\psi_{111} \psi_{212} = \psi_{211} \psi_{112}, \\
\psi_{121} \psi_{222} = \psi_{221} \psi_{122}, \\
\psi_{111} \psi_{122} = \psi_{112} \psi_{121}, \\
\psi_{211} \psi_{222} = \psi_{212} \psi_{221}, \label{BiseparableCAB5} \\
\psi_{211} \psi_{122} = \psi_{212} \psi_{121}. \label{BiseparableCAB6}
\end{align}
The bi-separability conditions of $A|BC$ and $B|CA$ can be derived by the same way. For complete separability, the three-qubit states would satisfy an extra condition besides equations (\ref{BiseparableCAB1}) to (\ref{BiseparableCAB6}), i.e.
\begin{equation}
\psi_{211} \psi_{122} = \psi_{212} \psi_{121} = \psi_{112} \psi_{221}.
\end{equation}

With respect to the core tensor of three qubits, we rewrite equations (\ref{BiseparableCAB1}) to (\ref{BiseparableCAB6}) as
\begin{align}
t_{111} t_{222} = t_{112} t_{221}, \label{CoreBiseparableCAB1} \\
t_{111} t_{212} = t_{211} t_{112}, \\
t_{121} t_{222} = t_{221} t_{122}, \\
t_{111} t_{122} = t_{112} t_{121}, \\
t_{211} t_{222} = t_{212} t_{221}, \label{CoreBiseparableCAB5} \\
t_{211} t_{122} = t_{212} t_{121}. \label{CoreBiseparableCAB6}
\end{align}
With HOSVD, we found that elements of the core tensor $\left[ t_{ijk} \right] \in \mathcal{T}_\psi$ have to satisfy only equation (\ref{CoreBiseparableCAB6}) to show the bi-separability of $C|AB$. In other words, we can derive equations (\ref{CoreBiseparableCAB1}) to (\ref{CoreBiseparableCAB5}) with equations (\ref{t111}), (\ref{t222}) and (\ref{CoreBiseparableCAB6}). Similarly, the following equations (\ref{CoreBiseparableABC6}) and (\ref{CoreBiseparableBCA6}) are the only condition to show the bi-separability of $A|BC$ and $B|CA$, respectively:-
\begin{align}
t_{112} t_{221} & = t_{212} t_{121}, \label{CoreBiseparableABC6} \\
t_{112} t_{221} & = t_{211} t_{122}. \label{CoreBiseparableBCA6}
\end{align}

Equations (\ref{ThreeQubitsFinalAllOrtho2a}), (\ref{ThreeQubitsFinalAllOrtho2b}) or (\ref{ThreeQubitsFinalAllOrtho2c}) inform us about the separability of three qubits. For bi-separable and completely separable states, these equations are automatically satisfied. Meanwhile, for genuinely entangled three-qubit states, equations (\ref{ThreeQubitsFinalAllOrtho2a}), (\ref{ThreeQubitsFinalAllOrtho2b}) or (\ref{ThreeQubitsFinalAllOrtho2c}) determine the linear dependency of the complex phases between $\bar{t}_{112} \bar{t}_{221} t_{122} t_{211}$, $\bar{t}_{121} \bar{t}_{212} t_{112} t_{221}$ and $\bar{t}_{122} \bar{t}_{211} t_{121} t_{212}$. However, since the complex phases will be canceled out in equations (\ref{ThreeQubitsFinalAllOrtho1a}) or (\ref{ThreeQubitsFinalAllOrtho1b}), their linear dependency is not important to us for the rest of the discussion.

\subsection{Special states of three qubits} \label{SubSecSpecialStatesThreeQubits}

Once we simultaneously diagonalized the one-body reduced density matrices of three qubits, there are three ways equations (\ref{ThreeQubitsFinalAllOrtho1a}) or (\ref{ThreeQubitsFinalAllOrtho1b}) can be satisfied. We explore all three possibilities in the following subsections and determine the special states in each case.

\subsubsection{Case 1: $\sigma_1^{(1)2} = \sigma_1^{(2)2} = \sigma_1^{(3)2}$}

Under this condition, equation (\ref{ThreeQubitsFinalAllOrtho1a}) is satisfied automatically. In addition, the following equations have to be satisfied:-
\begin{align}
\left| t_{121} \right|^2 + \left| t_{122} \right|^2 = \left| t_{211} \right|^2 + \left| t_{212} \right|^2, \\
\left| t_{112} \right|^2 + \left| t_{122} \right|^2 = \left| t_{211} \right|^2 + \left| t_{221} \right|^2.
\end{align}
If we set $\left| t_{112} \right|^2 = \left| t_{121} \right|^2 = \left| t_{122} \right|^2 = \left| t_{211} \right|^2 = \left| t_{212} \right|^2 = \left| t_{221} \right|^2 = 0$, we get the generalized GHZ states,
\begin{align}
\left| \text{GHZ} \right\rangle = t_{111} \left| 111 \right\rangle + t_{222} \left| 222 \right\rangle, \label{GHZ}
\end{align}
which is a special state under this condition.

With HOSVD, the completely separable states have the following one-body reduced density matrices
\begin{align}
\rho_d^A = \rho_d^B = \rho_d^C = \begin{pmatrix} 1 & 0 \\ 0 & 0 \end{pmatrix}.
\end{align}
Therefore, completely separable states are also under this condition.

\subsubsection{Case 2: Either $\sigma_1^{(1)2} = \sigma_1^{(2)2}$, $\sigma_1^{(1)2} = \sigma_1^{(3)2}$ or $\sigma_1^{(2)2} = \sigma_1^{(3)2}$}

For each of the possibilities, we list down the additional conditions that have to be satisfied and the respective special states.
\begin{enumerate}
\item{
If $\sigma_1^{(1)2} = \sigma_1^{(2)2}$, then it is necessary that $\left| t_{121} \right|^2 = \left| t_{211} \right|^2$ and $\left| t_{122} \right|^2 = \left| t_{212} \right|^2$. From equation (\ref{t111}), we have
\begin{equation*}
t_{111} = - \frac{\bar{t}_{221} (t_{121} t_{212} - t_{122} t_{211})}{t_{212} \bar{t}_{211} - t_{122} \bar{t}_{121}}.
\end{equation*}
Computing $\left| t_{111} \right|^2$, it is not difficult to show that
\begin{align*}
\left| t_{111} \right|^2 & = \left[ \frac{\bar{t}_{221} (t_{121} t_{212} - t_{122} t_{211})}{t_{212} \bar{t}_{211} - t_{122} \bar{t}_{121}} \right] \left[ \frac{t_{221} (\bar{t}_{121} \bar{t}_{212} - \bar{t}_{122} \bar{t}_{211})}{\bar{t}_{212} t_{211} - \bar{t}_{122} t_{121}}\right] \\ & = \left| t_{221} \right|^2.
\end{align*}
Similarly, we can show that $\left| t_{222} \right|^2 = \left| t_{112} \right|^2$. This implies that the $n$-mode singular values become
\begin{align*}
\sigma_1^{(1)2} & = \sigma_1^{(2)2} = \frac{1}{2}, \\
\sigma_2^{(1)2} & = \sigma_2^{(2)2} = \frac{1}{2}, \\
\sigma_1^{(3)2} & = 2 (\left| t_{111} \right|^2 + \left| t_{121} \right|^2), \\
\sigma_2^{(3)2} & = 2 (\left| t_{112} \right|^2 + \left| t_{122} \right|^2).
\end{align*}
This corresponds to states where $(\sigma_1^{(1)2}, \sigma_1^{(2)2}, \sigma_1^{(3)2}) = (\frac{1}{2}, \frac{1}{2}, \sigma_1^{(3)2})$.

A special state under this condition is when $\left| t_{121} \right|^2 = \left| t_{211} \right|^2 = \left| t_{122} \right|^2 = \left| t_{212} \right|^2 = 0$, i.e.
\begin{align}
\left| \text{S}_1 \right\rangle = t_{111} \left| 111 \right\rangle + t_{112} \left| 112 \right\rangle + t_{221} \left| 221 \right\rangle + t_{222} \left| 222 \right\rangle, & \label{Slice1} \\
\bar{t}_{111} t_{112} + \bar{t}_{221} t_{222} = 0. & \label{AllOrthoSlice1}
\end{align}
Equation (\ref{AllOrthoSlice1}) enables us to write
\begin{equation*}
\left| t_{221} \right|^2 = \frac{\left| t_{111} \right|^2 \left| t_{112} \right|^2}{\left| t_{222} \right|^2}.
\end{equation*}
Hence,
\begin{align*}
\sigma_1^{(3)2} & = \left| t_{111} \right|^2 + \frac{\left| t_{111} \right|^2 \left| t_{112} \right|^2}{\left| t_{222} \right|^2} \\ & = \frac{\left| t_{111} \right|^2}{\left| t_{222} \right|^2} (\left| t_{222} \right|^2 + \left| t_{112} \right|^2) \\ & = \frac{\left| t_{111} \right|^2}{\left| t_{222} \right|^2} \sigma_2^{(3)2}.
\end{align*}
Since $\sigma_1^{(3)2} + \sigma_2^{(3)2} = 1$, we find that equation above transforms into the followings:-
\begin{align*}
\sigma_1^{(3)2} & = \frac{\left| t_{111} \right|^2}{\left| t_{111} \right|^2 + \left| t_{222} \right|^2}, \\ \sigma_2^{(3)2} & = \frac{\left| t_{222} \right|^2}{\left| t_{111} \right|^2 + \left| t_{222} \right|^2}.
\end{align*}
From the ordering property of HOSVD, we have $\sigma_1^{(3)2} \geq \sigma_2^{(3)2}$, which leads us to conclude that $\left| t_{111} \right|^2 \geq \left| t_{222} \right|^2$. Now,
\begin{align*}
\sigma_1^{(3)2} - \sigma_1^{(1)2} & = \left| t_{221} \right|^2 - \left| t_{112} \right|^2 \\ & = \frac{\left| t_{111} \right|^2 \left| t_{112} \right|^2}{\left| t_{222} \right|^2} - \left| t_{112} \right|^2 \\ & = \frac{\left| t_{112} \right|^2}{\left| t_{222} \right|^2} (\left| t_{111} \right|^2 - \left| t_{222} \right|^2).
\end{align*}
Since $\sigma_1^{(3)2} \neq \sigma_1^{(1)2}$, we conclude that $\sigma_1^{(3)2} > \sigma_1^{(1)2}$.
}

\item{
If $\sigma_1^{(1)2} = \sigma_1^{(3)2}$, then it is necessary that $\left| t_{112} \right|^2 = \left| t_{211} \right|^2$ and $\left| t_{122} \right|^2 = \left| t_{221} \right|^2$. Using a similar proof from above, this corresponds to states where $(\sigma_1^{(1)2}, \sigma_1^{(2)2}, \sigma_1^{(3)2}) = (\frac{1}{2}, \sigma_1^{(2)2}, \frac{1}{2})$.

A special state under this condition is when $\left| t_{112} \right|^2 = \left| t_{211} \right|^2 = \left| t_{122} \right|^2 = \left| t_{221} \right|^2 = 0$, i.e.
\begin{align}
\left| \text{S}_2 \right\rangle = t_{111} \left| 111 \right\rangle + t_{121} \left| 121 \right\rangle + t_{212} \left| 212 \right\rangle + t_{222} \left| 222 \right\rangle, & \label{Slice2} \\
\bar{t}_{111} t_{121} + \bar{t}_{212} t_{222} = 0. &
\end{align}
Here, $\sigma_1^{(2)2} > \sigma_1^{(1)2}$.
}

\item{
If $\sigma_1^{(2)2} = \sigma_1^{(3)2}$, then it is necessary that $\left| t_{112} \right|^2 = \left| t_{121} \right|^2$ and $\left| t_{212} \right|^2 = \left| t_{221} \right|^2$. Using a similar proof from above, this corresponds to states where $(\sigma_1^{(1)2}, \sigma_1^{(2)2}, \sigma_1^{(3)2}) = (\sigma_1^{(1)2}, \frac{1}{2}, \frac{1}{2})$.

A special state under this condition is when $\left| t_{112} \right|^2 = \left| t_{121} \right|^2 = \left| t_{212} \right|^2 = \left| t_{221} \right|^2 = 0$, i.e.
\begin{align}
\left| \text{S}_3 \right\rangle = t_{111} \left| 111 \right\rangle + t_{122} \left| 122 \right\rangle + t_{211} \left| 211 \right\rangle + t_{222} \left| 222 \right\rangle, & \label{Slice3} \\
\bar{t}_{111} t_{211} + \bar{t}_{122} t_{222} = 0. &
\end{align}
Here, $\sigma_1^{(1)2} > \sigma_1^{(2)2}$.
}
\end{enumerate}

We also note that the bi-separable states fall under this case. For example, the HOSVD of the bi-separable state $C|AB$ has the following one-body reduced density matrices:-
\begin{align}
\rho_d^A = \rho_d^B & = \begin{pmatrix} \left| t_{111} \right|^2 & 0 \\ 0 & \left| t_{221} \right|^2 \end{pmatrix}, \\
\rho_d^C & = \begin{pmatrix} 1 & 0 \\ 0 & 0 \end{pmatrix}.
\end{align}

\subsubsection{Case 3: $\sigma_1^{(1)2} \neq \sigma_1^{(2)2} \neq \sigma_1^{(3)2}$}

Since there is no special requirements on the $n$-mode singular values ($\sigma_1^{(n)2}$ where $n=1,2,3$), this is where a generic genuinely entangled three-qubit state would be located. By rearranging the terms in equation (\ref{ThreeQubitsFinalAllOrtho1a}), we arrive at the following form:-
\begin{align}
a \sigma_1^{(1)2} + b \sigma_1^{(2)2} + c \sigma_1^{(3)2} = 0, \label{ThreeQubitsFinalAllOrthoTransform}
\end{align}
where $a = \left| t_{112} \right|^2 - \left| t_{121} \right|^2$, $b = \left| t_{211} \right|^2 - \left| t_{112} \right|^2$ and $c = \left| t_{121} \right|^2 - \left| t_{211} \right|^2$. This is an equation of a plane that cuts through the origin with normal vector $\vec{n} = (a,b,c)$ and an extra condition of $a + b + c = 0$. Without loss of generality, we consider the case when $c = - (a + b)$, with $a$ and $b$ being positive. Equation (\ref{ThreeQubitsFinalAllOrthoTransform}) will then become
\begin{align}
& a (\sigma_1^{(1)2} - \sigma_1^{(3)2}) + b (\sigma_1^{(2)2} - \sigma_1^{(3)2}) = 0 \nonumber \\
\Rightarrow & a (\sigma_1^{(1)2} + \sigma_1^{(2)2} - \sigma_1^{(3)2}) + b (\sigma_1^{(1)2} + \sigma_1^{(2)2} - \sigma_1^{(3)2}) = a \sigma_1^{(2)2} + b \sigma_1^{(1)2} \nonumber \\
\Rightarrow & \sigma_1^{(1)2} + \sigma_1^{(2)2} - \sigma_1^{(3)2} = \frac{a}{a+b} \sigma_1^{(2)2} + \frac{b}{a+b} \sigma_1^{(1)2}. \label{ThreeQubitsPolytope}
\end{align}
Equation (\ref{ThreeQubitsPolytope}) shows that the sum $\sigma_1^{(1)2} + \sigma_1^{(2)2} - \sigma_1^{(3)2}$ is a convex combination of $\sigma_1^{(1)2}$ and $\sigma_1^{(2)2}$. Since $0.5 \leq \sigma_1^{(1)2}, \sigma_1^{(2)2} \leq 1$, the upper bound of equation (\ref{ThreeQubitsPolytope}) is therefore 1, i.e.
\begin{equation}
\sigma_1^{(1)2} + \sigma_1^{(2)2} - \sigma_1^{(3)2} \leq 1. \label{ThreeQubitsPolytopeBounded1}
\end{equation}
Similar argument can be carried out for $b = - (a + c)$ and $a = - (b + c)$, leading us to
\begin{align}
\sigma_1^{(1)2} + \sigma_1^{(3)2} - \sigma_1^{(2)2} \leq 1, \label{ThreeQubitsPolytopeBounded2} \\
\sigma_1^{(2)2} + \sigma_1^{(3)2} - \sigma_1^{(1)2} \leq 1. \label{ThreeQubitsPolytopeBounded3}
\end{align}

If we let $\left| t_{112} \right|^2 = \left| t_{121} \right|^2 = \left| t_{211} \right|^2 = 0$, then $\left| t_{222} \right|^2$ has to be zero as well due to the all-orthogonality conditions (\ref{ThreeQubitsAllOrthogonality1}), (\ref{ThreeQubitsAllOrthogonality2}) and (\ref{ThreeQubitsAllOrthogonality3}). Similarly, we can let $\left| t_{122} \right|^2 = \left| t_{212} \right|^2 = \left| t_{221} \right|^2 = 0$ and $\left| t_{111} \right|^2$ is automatically zero. We will then have the following equivalent special states,
\begin{align}
\left| \text{B}_1 \right\rangle = t_{111} \left| 111 \right\rangle + t_{122} \left| 122 \right\rangle + t_{212} \left| 212 \right\rangle + t_{221} \left| 221 \right\rangle, \label{Beechnut1} \\
\left| \text{B}_2 \right\rangle = t_{112} \left| 112 \right\rangle + t_{121} \left| 121 \right\rangle + t_{211} \left| 211 \right\rangle + t_{222} \left| 222 \right\rangle. \label{Beechnut2}
\end{align}

\subsection{The polytope of three qubits}

Due to the momentum map that identifies multipartite states with its one-body reduced density matrices, the critical points of the total variance function of the multipartite states are equivalent to the critical spectra of the one-body reduced density matrices, which can then be used to parametrize the stochastic local operation and classical communication (SLOCC) classes of multipartite entanglement \cite{Sawicki2012}. In \cite{Walter2013,Sawicki2014}, the authors showed the construction of the entanglement polytope of three qubits by SLOCC. Here, we show that it is also possible to construct a polytope of three qubits via the $n$-mode singular values classified by the local unitary (LU) operations through HOSVD. Table \ref{tableHOSVD} summarizes our findings in Section \ref{SubSecSpecialStatesThreeQubits} according to the behavior of the largest eigenvalue ($\sigma_1^{(n)2}$ where $n=1,2,3$) of one-body reduced density matrices of three qubits.

\begin{figure}[H]
\begin{center}
\begin{tabular}{c c}
\includegraphics[width=6cm]{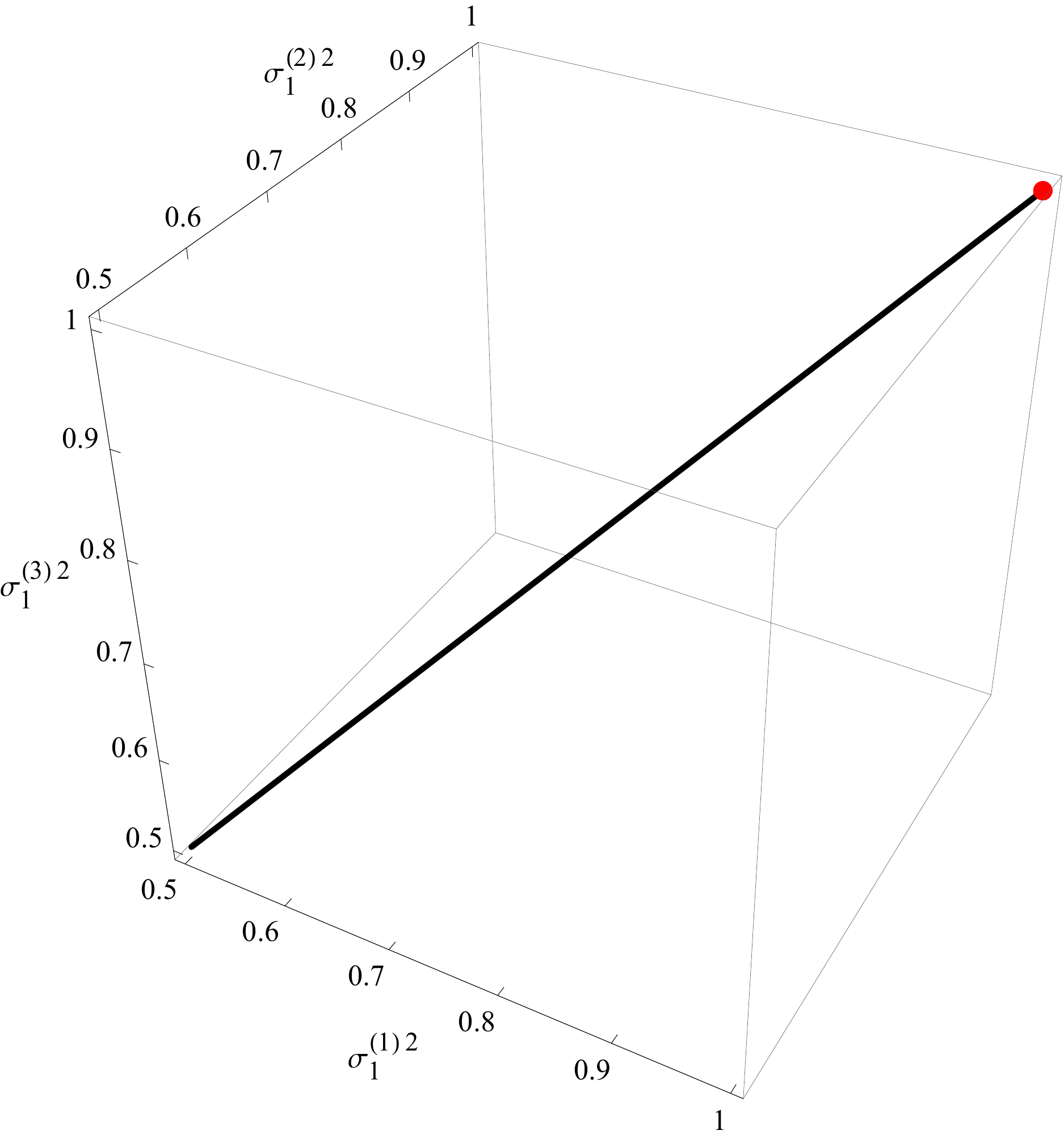} & \includegraphics[width=6cm]{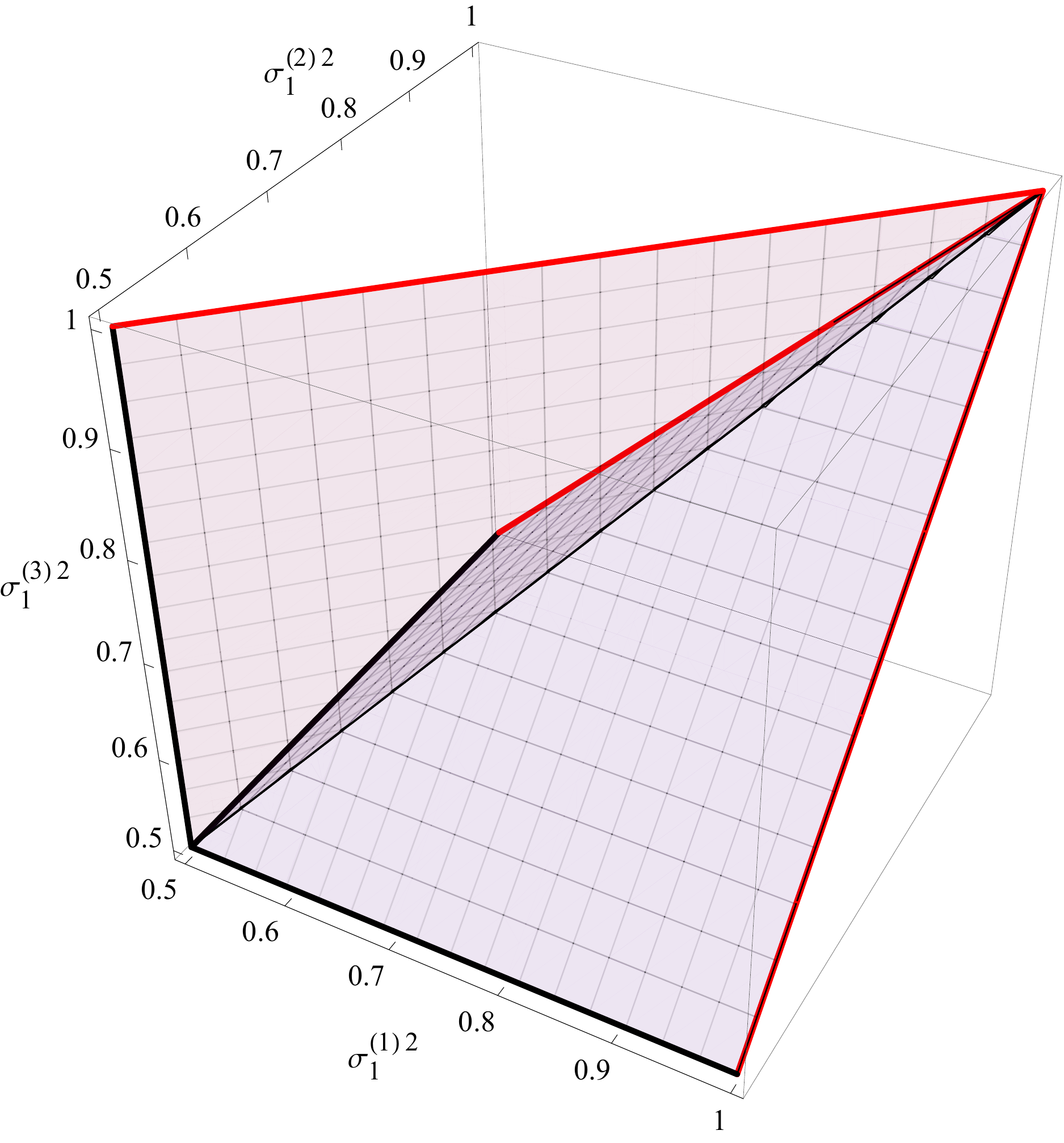} \\ (a) & (b) \\ \includegraphics[width=6cm]{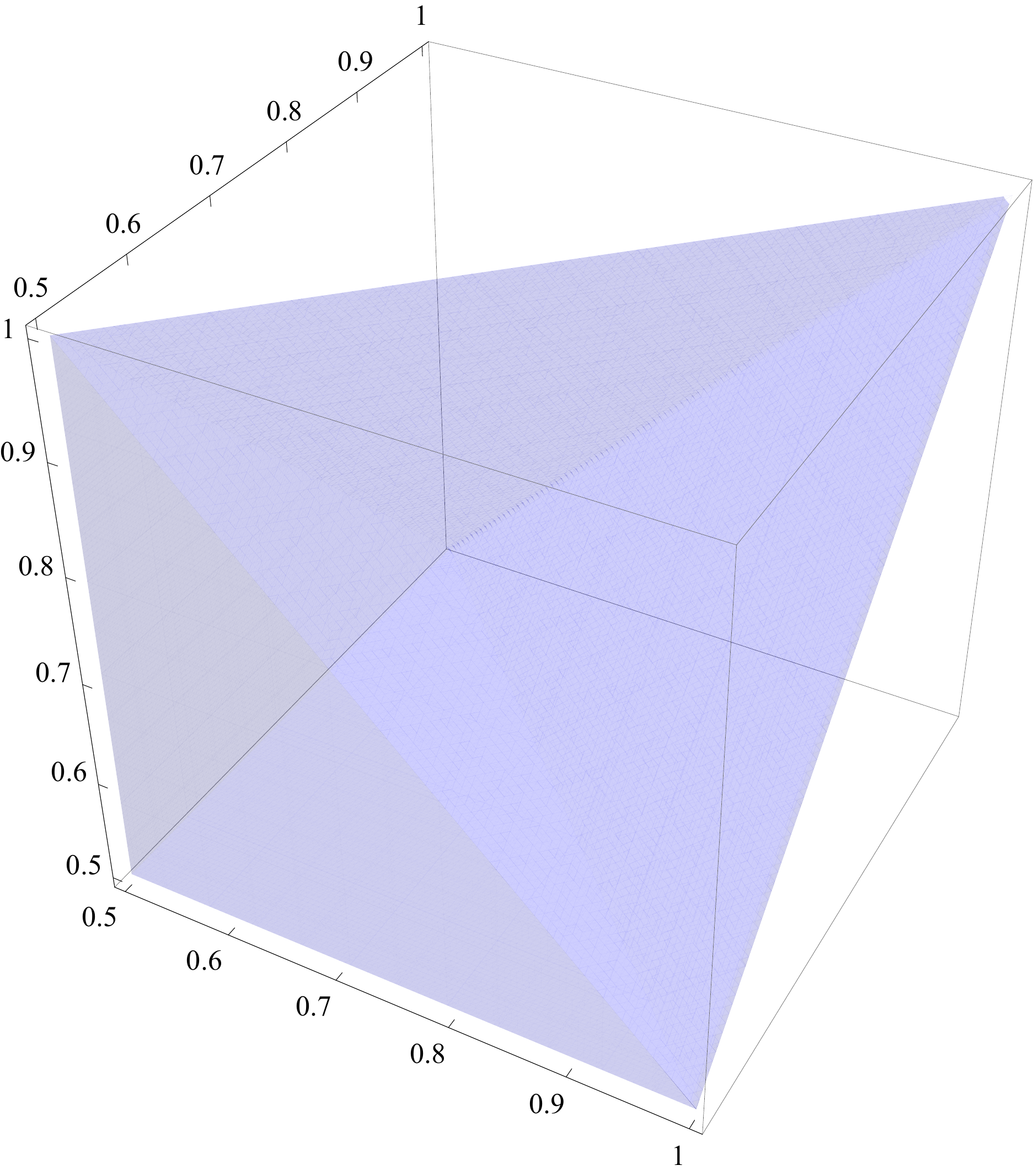} & \includegraphics[width=6cm]{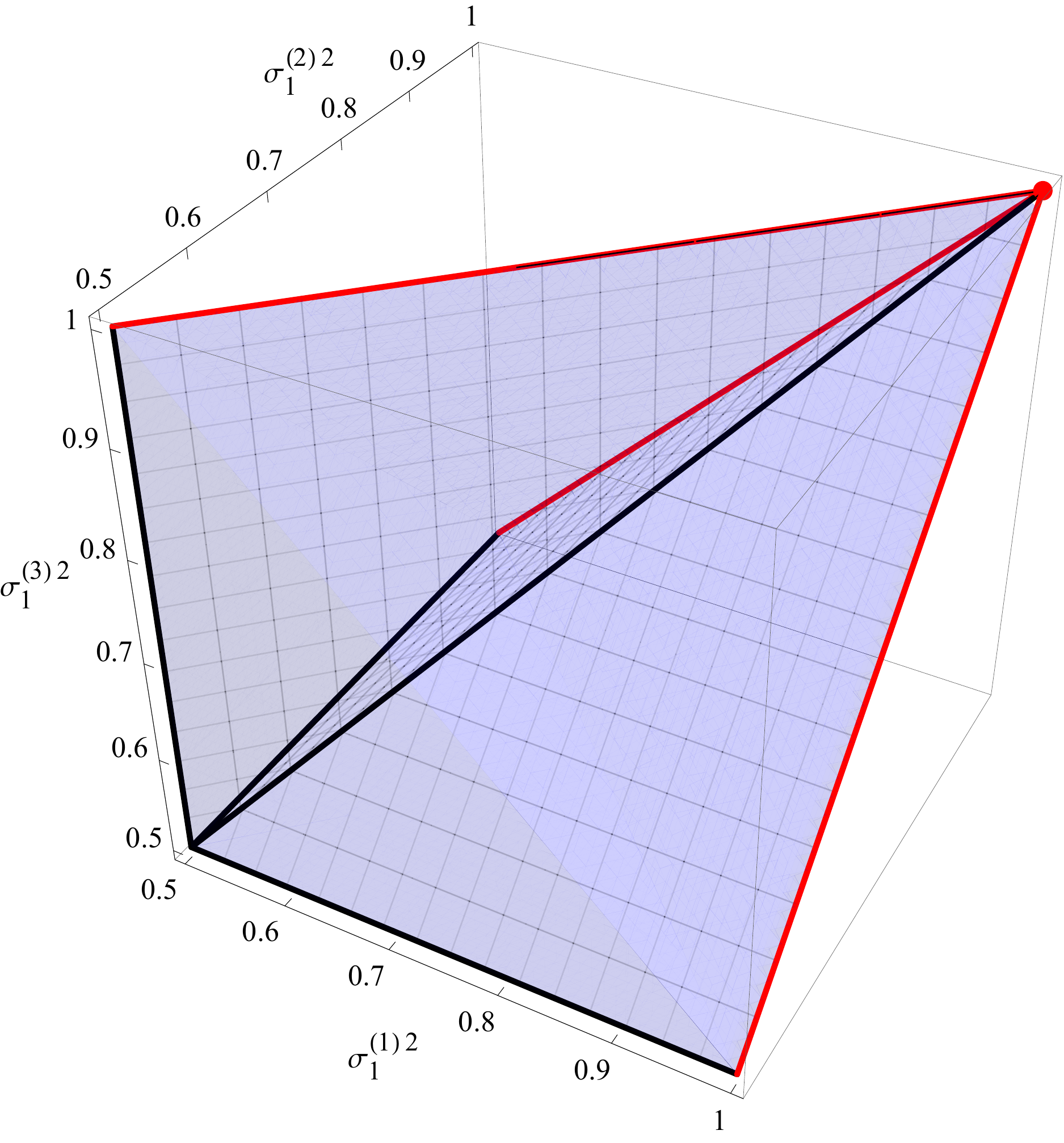} \\ (c) & (d) \\
\end{tabular}
\end{center}
\caption{(a) Case 1: The black line represents states when $\sigma_1^{(1)2} = \sigma_1^{(2)2} = \sigma_1^{(3)2}$, while the red point is the completely separable states. (b) Case 2: The black lines represent states of the form $(\sigma_1^{(1)2}, \frac{1}{2}, \frac{1}{2})$, $(\frac{1}{2}, \sigma_1^{(2)2}, \frac{1}{2})$ and $(\frac{1}{2}, \frac{1}{2}, \sigma_1^{(3)2})$, while the red lines are the bi-separable states. The mesh planes are $\left| \text{S}_1 \right\rangle$, $\left| \text{S}_2 \right\rangle$ and $\left| \text{S}_3 \right\rangle$. (c) Case 3: The whole polytope bounded by the three inequalities (\ref{ThreeQubitsPolytopeBounded1}) to (\ref{ThreeQubitsPolytopeBounded3}). (d) The combination of all cases.} \label{EntanglementPolytope}
\end{figure}

As mentioned in Section \ref{SubSecHOSVDThreeQubits}, due to the ordering property of higher order singular value decomposition (HOSVD), $\sigma_1^{(n)2} \geq \sigma_2^{(n)2}$ for $n = 1, 2, 3$. Therefore, $0.5 \leq \sigma_1^{(n)2} \leq 1$. Combining with other constraints discussed in Section \ref{SubSecSpecialStatesThreeQubits} and summarized in Table \ref{tableHOSVD}, we plot the polytope of three qubits by LU operations in Figure \ref{EntanglementPolytope}. We find that the polytope perfectly includes all the special states that we discovered by HOSVD.

\begin{table}[H]
\scriptsize{
\begin{center}
\begin{tabular}{|l|l|}
\hline
Case & States \\
\hline
1. $\sigma_1^{(1)2} = \sigma_1^{(2)2} = \sigma_1^{(3)2}$ & (a) General states, \\ & $\quad \enspace \left| t_{121} \right|^2 + \left| t_{122} \right|^2 = \left| t_{211} \right|^2 + \left| t_{212} \right|^2,$ \\ & $\quad \enspace \left| t_{112} \right|^2 + \left| t_{122} \right|^2 = \left| t_{211} \right|^2 + \left| t_{221} \right|^2.$ \\
\cline{2-2}
& (b) Generalized GHZ states, \\ & $\quad \enspace \left| \text{GHZ} \right\rangle = t_{111} \left| 111 \right\rangle + t_{222} \left| 222 \right\rangle.$ \\
\cline{2-2}
& (c) Completely separable states, \\ & $\quad \enspace \left| \text{Sep} \right\rangle = t_{111} \left| 111 \right\rangle.$ \\
\hline
2. $\sigma_1^{(1)2} = \sigma_1^{(2)2}$, $\sigma_1^{(1)2} = \sigma_1^{(3)2}$ & (a) (1) $\sigma_1^{(1)2} = \sigma_1^{(2)2}$, $(\sigma_1^{(1)2}, \sigma_1^{(2)2}, \sigma_1^{(3)2}) = (\frac{1}{2}, \frac{1}{2}, \sigma_1^{(3)2}),$ \\ \quad or $\sigma_1^{(2)2} = \sigma_1^{(3)2}$ & $\qquad \quad \left| t_{121} \right|^2 = \left| t_{211} \right|^2,\, \left| t_{122} \right|^2 = \left| t_{212} \right|^2,$ \\ & $\qquad \quad \left| t_{111} \right|^2 = \left| t_{221} \right|^2,\, \left| t_{112} \right|^2 = \left| t_{222} \right|^2.$ \\
& \phantom{(a)} (2) $\sigma_1^{(1)2} = \sigma_1^{(3)2}$, $(\sigma_1^{(1)2}, \sigma_1^{(2)2}, \sigma_1^{(3)2}) = (\frac{1}{2}, \sigma_1^{(2)2}, \frac{1}{2}),$ \\ & $\qquad \quad \left| t_{112} \right|^2 = \left| t_{211} \right|^2,\, \left| t_{122} \right|^2 = \left| t_{221} \right|^2,$ \\ & $\qquad \quad \left| t_{111} \right|^2 = \left| t_{212} \right|^2,\, \left| t_{121} \right|^2 = \left| t_{222} \right|^2.$ \\
& \phantom{(a)} (3) $\sigma_1^{(2)2} = \sigma_1^{(3)2}$, $(\sigma_1^{(1)2}, \sigma_1^{(2)2}, \sigma_1^{(3)2}) = (\sigma_1^{(1)2}, \frac{1}{2}, \frac{1}{2}),$ \\ & $\qquad \quad \left| t_{112} \right|^2 = \left| t_{121} \right|^2,\, \left| t_{212} \right|^2 = \left| t_{221} \right|^2,$ \\ & $\qquad \quad \left| t_{111} \right|^2 = \left| t_{122} \right|^2,\, \left| t_{211} \right|^2 = \left| t_{222} \right|^2.$ \\
\cline{2-2}
& (b) (1) $\sigma_1^{(3)2} > \sigma_1^{(1)2} = \sigma_1^{(2)2},$ \\ & $\qquad \quad \left| \text{S}_1 \right\rangle = t_{111} \left| 111 \right\rangle + t_{112} \left| 112 \right\rangle + t_{221} \left| 221 \right\rangle + t_{222} \left| 222 \right\rangle,$ \\ & $\qquad \quad \bar{t}_{111} t_{112} + \bar{t}_{221} t_{222} = 0.$ \\
& \phantom{(b)} (2) $\sigma_1^{(2)2} > \sigma_1^{(1)2} = \sigma_1^{(3)2},$ \\ & $\qquad \quad \left| \text{S}_2 \right\rangle = t_{111} \left| 111 \right\rangle + t_{121} \left| 121 \right\rangle + t_{212} \left| 212 \right\rangle + t_{222} \left| 222 \right\rangle,$ \\ & $\qquad \quad \bar{t}_{111} t_{121} + \bar{t}_{212} t_{222} = 0.$ \\
& \phantom{(b)} (3) $\sigma_1^{(1)2} > \sigma_1^{(2)2} = \sigma_1^{(3)2},$ \\ & $\qquad \quad \left| \text{S}_3 \right\rangle = t_{111} \left| 111 \right\rangle + t_{122} \left| 122 \right\rangle + t_{211} \left| 211 \right\rangle + t_{222} \left| 222 \right\rangle,$ \\ & $\qquad \quad \bar{t}_{111} t_{211} + \bar{t}_{122} t_{222} = 0.$ \\
\cline{2-2}
& (c) (1) Bi-separable state $C|AB$, \\ & $\qquad \quad \left| \text{Bi-Sep}_{C|AB} \right\rangle = t_{111} \left| 111 \right\rangle + t_{221} \left| 221 \right\rangle.$ \\
& \phantom{(c)} (2) Bi-separable state $B|CA$, \\ & $\qquad \quad \left| \text{Bi-Sep}_{B|CA} \right\rangle = t_{111} \left| 111 \right\rangle + t_{212} \left| 212 \right\rangle.$ \\
& \phantom{(c)} (3) Bi-separable state $A|BC$, \\ & $\qquad \quad \left| \text{Bi-Sep}_{A|BC} \right\rangle = t_{111} \left| 111 \right\rangle + t_{122} \left| 122 \right\rangle.$ \\
\hline
3. $\sigma_1^{(1)2} \neq \sigma_1^{(2)2} \neq \sigma_1^{(3)2}$ & (a) General states, \\ & \phantom{(a)} $\left| t_{112} \right|^2 \left[ \sigma_1^{(1)2} - \sigma_1^{(2)2} \right] + \left| t_{211} \right|^2 \left[ \sigma_1^{(2)2} - \sigma_1^{(3)2} \right]$ \\ & $\qquad + \left| t_{121} \right|^2 \left[ \sigma_1^{(3)2} - \sigma_1^{(1)2} \right] = 0,$ \\
& \phantom{(a)} $\bar{t}_{112} \bar{t}_{221} (t_{122} t_{211} - t_{121} t_{212}) + \bar{t}_{121} \bar{t}_{212} (t_{112} t_{221} - $ \\ & $\qquad t_{122} t_{211}) + \bar{t}_{122} \bar{t}_{211} (t_{121} t_{212} - t_{112} t_{221}) = 0;$ \\
& \phantom{(a)} $\sigma_1^{(1)2} + \sigma_1^{(2)2} - \sigma_1^{(3)2} \leq 1,$ \\ & \phantom{(a)} $\sigma_1^{(1)2} + \sigma_1^{(3)2} - \sigma_1^{(2)2} \leq 1,$ \\ & \phantom{(a)} $\sigma_1^{(2)2} + \sigma_1^{(3)2} - \sigma_1^{(1)2} \leq 1.$ \\
\cline{2-2}
& (b) $\left| \text{B}_1 \right\rangle = t_{111} \left| 111 \right\rangle + t_{122} \left| 122 \right\rangle + t_{212} \left| 212 \right\rangle + t_{221} \left| 221 \right\rangle,$ \\ & \phantom{(b)} $\left| \text{B}_2 \right\rangle = t_{112} \left| 112 \right\rangle + t_{121} \left| 121 \right\rangle + t_{211} \left| 211 \right\rangle + t_{222} \left| 222 \right\rangle.$ \\
\hline
\end{tabular}
\end{center}
\caption{HOSVD of three qubits and their respective cases} \label{tableHOSVD}
}
\end{table}

\subsection{The exceptional states of three qubits} \label{SubSecExceptionalStates}

In this subsection, we would like to make a comparison between our results and that in \cite{Carteret2000}. We summarize their classification in Table \ref{tableCarteretI}. Since it was shown that exceptional states have enlarged stabilizers, we choose to include the stabilizers in the table.

In Table \ref{tableCarteretII}, we check each of the exceptional states in \cite{Carteret2000} to see if they are in the higher order singular value decomposition (HOSVD) form, and their respective one-body reduced density matrices. Of all the exceptional states, only Slice states are not in the HOSVD form. Therefore, we compute the eigenvalues of its one-body reduced density matrices in Table \ref{tableCarteretII}. In addition, we note that the generic states in \cite{Carteret2000} are not exceptional states, since their stabilizers are discrete.

By comparison, it is not difficult to see that the set of special states ($\left| \text{S}_1 \right\rangle$, $\left| \text{S}_2 \right\rangle$, $\left| \text{S}_3 \right\rangle$) correspond to the Slice states in \cite{Carteret2000}, while ($\left| \text{B}_1 \right\rangle$, $\left| \text{B}_2 \right\rangle$) corresponds to the Beechnut states. Because of this correspondence, we choose to label those special states as they are.

\begin{table}[H]
\scriptsize{
\begin{center}
\begin{tabular}{|l|l|}
\hline
Local unitary classes & Exceptional states and their stabilizers \\
\hline
1. Generic states & Generic three-qubit states \\ & $\text{Stab} = \left\{ (e^{i \varphi}, U, V, W) = (1, \textbf{1}, \textbf{1}, \textbf{1})\right\}$ \\
\hline
2. Bystander's states & (a) $T_2$ is singular, i.e. $\left| \psi \right\rangle = \psi_{211} \left| 211 \right\rangle$ \\ & \phantom{(a)} $\text{Stab} = \left\{ (e^{i \varphi}, U, V, W) = \left[e^{i \varphi}, \begin{pmatrix} e^{i (\varphi + \gamma + \eta)} & 0 \\ 0 & e^{-i (\varphi + \gamma + \eta)} \end{pmatrix},\right.\right.$ \\ & \phantom{(a)} \hspace{35pt} $\left.\left. \begin{pmatrix} e^{i \gamma} & 0 \\ 0 & e^{-i \gamma} \end{pmatrix}, \begin{pmatrix} e^{i \eta} & 0 \\ 0 & e^{-i \eta} \end{pmatrix}\right] \right\}$ \\
\cline{2-2}
& (b) $T_2$ is not singular, i.e. $\left| \psi \right\rangle = \psi_{211} \left| 211 \right\rangle + \psi_{222} \left| 222 \right\rangle$ \\ & \phantom{(a)} (i) If $\left| \psi_{211} \right| \neq \left| \psi_{222} \right|$ \\ & \phantom{(a) (i)} $\text{Stab} = \left\{ (e^{i \varphi}, U, V, W) = \left[e^{i \varphi}, \begin{pmatrix} e^{i \varphi} & 0 \\ 0 & e^{-i \varphi} \end{pmatrix},\right.\right.$ \\ & \phantom{(a) (i)} \hspace{35pt} $\left.\left. \begin{pmatrix} e^{i \gamma} & 0 \\ 0 & e^{-i \gamma} \end{pmatrix}, \begin{pmatrix} e^{-i \gamma} & 0 \\ 0 & e^{i \gamma} \end{pmatrix}\right] \right\}$ \\ & \phantom{(a)} (ii) If $\left| \psi_{211} \right| = \left| \psi_{222} \right|$ \\ & \phantom{(a) (ii)} $\text{Stab} = \left\{ (e^{i \varphi}, U, V, W) = \left[e^{i \varphi}, \begin{pmatrix} e^{i \varphi} & 0 \\ 0 & e^{-i \varphi} \end{pmatrix},\right.\right.$ \\ & \phantom{(a) (ii)} \hspace{35pt} $\left.\left. \begin{pmatrix} v_{11} & v_{12} \\ -\bar{v}_{12} & \bar{v}_{11} \end{pmatrix}, \bar{V} \right] \right\}$ \\
\hline
3. Slice states & States of the form $\left| \psi \right\rangle = p \left| 111 \right\rangle + bc \left| 221 \right\rangle + bd \left| 222 \right\rangle$ and its \\ & qubit-relabeling permutations \\ & (a) If $u_{12} = 0$, \\ & \phantom{(a)} $\text{Stab} = \left\{ (e^{i \varphi}, U, V, W) = \left[\epsilon_1 1, e^{i \theta \sigma_3}, \epsilon_2 e^{-i \theta \sigma_3}, \epsilon_1 \epsilon_2 \textbf{1} \right] \right\}$ \\ & (b) If $u_{11} = 0$, then $p^2 = \left| b \right|^2 (\left| c \right|^2 + \left| d \right|^2)$ \\ & \phantom{(b)} $\text{Stab} = \left\{ (e^{i \varphi}, U, V, W) = \left[\epsilon_1 i, \begin{pmatrix} 0 & e^{i \theta} \\ -e^{-i \theta} & 0 \end{pmatrix},\right.\right.$ \\ & \phantom{(b)} \hspace{5pt} $\left.\left. \epsilon_2 \begin{pmatrix} 0 & e^{-i (\theta + \chi)} \\ -e^{i (\theta + \chi)} & 0 \end{pmatrix}, \epsilon_1 \epsilon_2 \begin{pmatrix} -i \frac{\left| bc \right|}{p} & -i e^{i \chi} \frac{\bar{b} \bar{d}}{p} \\ -i e^{-i \chi} \frac{bd}{p} & i \frac{\left| bc \right|}{p} \end{pmatrix} \right] \right\}$ \\ & \phantom{(b)} where $p$ is real, $b, c, d$ are complex, $\epsilon_1, \epsilon_2 = \pm 1$, and \\ & \phantom{(b)} $\chi = \arg{(bc)}$. \\
\hline
4. GHZ states & States of the form $\left| \psi \right\rangle = p \left| 111 \right\rangle + q \left| 222 \right\rangle$ \\ & (a) If $u_{12} = 0$, \\ & \phantom{(a)} $\text{Stab} = \left\{ (e^{i \varphi}, U, V, W) = \left[\epsilon_1 1, e^{i \theta \sigma_3}, e^{i \alpha \sigma_3}, e^{i \beta \sigma_3} \right] \right\}$ \\ & \phantom{(a)} where $\theta + \alpha + \beta = 0 (\text{mod } \pi)$. \\ & (b) If $\left| q \right| = p$, \\ & \phantom{(b)} $\text{Stab} = \left\{ (e^{i \varphi}, U, V, W) = \left[\epsilon_1 i, \begin{pmatrix} 0 & e^{i \theta} \\ -e^{-i \theta} & 0 \end{pmatrix},\right.\right.$ \\ & \phantom{(b)} \hspace{35pt} $\left.\left. \begin{pmatrix} 0 & e^{i \alpha} \\ -e^{-i \alpha} & 0 \end{pmatrix}, \begin{pmatrix} 0 & e^{i \beta} \\ -e^{-i \beta} & 0 \end{pmatrix} \right] \right\}$ \\ & \phantom{(b)} where $\theta + \alpha + \beta = \frac{\pi}{2} (\text{mod } \pi)$. \\
\hline
5. Beechnut states & States of the following forms:- \\ & $\left| \psi \right\rangle = wc \left| 111 \right\rangle + b \left| 212 \right\rangle + c \left| 221 \right\rangle$ \\ & $\left| \psi \right\rangle = wc \left| 112 \right\rangle + b \left| 211 \right\rangle + c \left| 222 \right\rangle$ \\ & $\text{Stab} = \left\{ (e^{i \varphi}, U, V, W) = \left[e^{i \varphi}, e^{i \varphi \sigma_3}, e^{i \varphi \sigma_3}, e^{-i \varphi \sigma_3} \right] \right\}$ \\
\hline
\end{tabular}
\end{center}
}
\caption{Local unitary classification of three qubits \cite{Carteret2000} and the stabilizers of respective exceptional states.} \label{tableCarteretI}
\end{table}

\begin{table}[H]
\scriptsize{
\begin{center}
\begin{tabular}{|l|l|l|}
\hline
Local unitary classes & HOSVD form & One-body reduced density matrices \\
\hline
1. Generic states & No & $\rho^A, \rho^B, \rho^C$ are generic one-body \\ & & reduced density matrices of three \\ & & qubits \\
\hline
2. Bystander's states (a) & Yes & $\rho^A = \begin{pmatrix} 0 & 0 \\ 0 & 1 \end{pmatrix}, \rho^B = \rho^C = \begin{pmatrix} 1 & 0 \\ 0 & 0 \end{pmatrix}$ \\
\phantom{2. }Bystander's states (b) (i) & Yes & $\rho^A = \begin{pmatrix} 0 & 0 \\ 0 & 1 \end{pmatrix},$\\ & & $\rho^B = \rho^C = \begin{pmatrix} \left| \psi_{211} \right|^2 & 0 \\ 0 & \left| \psi_{222} \right|^2 \end{pmatrix}$ \\
\phantom{2. }Bystander's states (b) (ii) & Yes & $\rho^A = \begin{pmatrix} 0 & 0 \\ 0 & 1 \end{pmatrix}, \rho^B = \rho^C = \begin{pmatrix} \frac{1}{2} & 0 \\ 0 & \frac{1}{2} \end{pmatrix}$ \\
\hline
3. Slice states & No & $\rho^A = \rho^B = \begin{pmatrix} p^2 & 0 \\ 0 & \left| b \right|^2 (\left| c \right|^2 + \left| d \right|^2) \end{pmatrix},$ \\ & & $\rho^C = \begin{pmatrix} p^2 + \left| bc \right|^2 & \left| b \right|^2 c \bar{d} \\ \left| b \right|^2 \bar{c} d & \left| bd \right|^2 \end{pmatrix}$ \\ & & Eigenvalues of $\rho^C$ are $\frac{1 \pm \sqrt{1 - 4 p^2 \left| bd \right|^2}}{2}$.\\ & & Any qubit-relabeling permutation \\ & & results in the same set of reduced \\ & & density matrices. \\
\hline
4. GHZ states & Yes & $\rho^A = \rho^B = \rho^C = \begin{pmatrix} p^2 & 0 \\ 0 & \left| q \right|^2 \end{pmatrix}$ \\
\hline
5. Beechnut states & Yes & $\rho^A = \begin{pmatrix} \left| wc \right|^2 & 0 \\ 0 & \left| b \right|^2 + \left| c \right|^2 \end{pmatrix},$ \\ & & $\rho^B = \begin{pmatrix} \left| wc \right|^2 + \left| b \right|^2 & 0 \\ 0 & \left| c \right|^2 \end{pmatrix},$ \\ & & $\rho^C = \begin{pmatrix} \left| wc \right|^2 + \left| c \right|^2 & 0 \\ 0 & \left| b \right|^2 \end{pmatrix}$ or \\ & & $\rho^C = \begin{pmatrix} \left| b \right|^2 & 0 \\ 0 & \left| wc \right|^2 + \left| c \right|^2 \end{pmatrix}$ \\
\hline
\end{tabular}
\end{center}
}
\caption{Local unitary classification of three qubits \cite{Carteret2000} and its one-body reduced density matrices.} \label{tableCarteretII}
\end{table}

\section{Conclusion}

From the all-orthogonality conditions of higher order singular value decomposition (HOSVD) for three qubits, we derived equation (\ref{ThreeQubitsFinalAllOrtho1a}) or equivalently equation (\ref{ThreeQubitsFinalAllOrtho1b}) that the $n$-mode singular values have to satisfy. We studied all possible scenarios that satisfy equation (\ref{ThreeQubitsFinalAllOrtho1a}) and computed all the special states of three qubits. Algebraically, the special states of three qubits are special zeroes of the polynomial (\ref{ThreeQubitsFinalAllOrtho1a}). The correspondence between the special states in our work with the exceptional states found in \cite{Carteret2000} shows that we recovered the LU classification of three qubits by using HOSVD.

As we have shown in Section \ref{SecHOSVD}, HOSVD simultaneously diagonalizes three-qubit states through LU actions, therefore our results are with respect to the LU equivalence. In comparison to the entanglement polytope constructed in \cite{Walter2013} which is based on the stochastic local operation and classical communication (SLOCC) equivalence, we did not recover the inequality
\begin{equation*}
\sigma_1^{(1)2} + \sigma_1^{(2)2} + \sigma_1^{(3)2} \leq 2.
\end{equation*}
This inequality separates the GHZ- and W-polytope in \cite{Walter2013}. It will be an interesting problem to check if HOSVD can be used to classify multipartite states by SLOCC. Technically, HOSVD can be used to characterize LU entanglement classes of multipartite states with more than three subsystems (for example, four qubits) or of higher local dimensions (for example, three qutrits). However, due to the multiplicative nature of tensor product, the complexity of the computation will increase exponentially. Another future problem that can be tackled is to simplify such calculation.

In summary, we studied explicitly the matrix unfolding and HOSVD \cite{Lathauwer2000} for three qubits. We showed that the matrix unfoldings of three qubits are related to their reduced density matrices, while HOSVD simultaneously diagonalizes the one-body reduced density matrices of three qubits. Due to the all-orthogonality conditions from HOSVD, we identified the special states of three qubits. Since the special states are in correspondence to the exceptional states \cite{Carteret2000}, we completely classified three-qubit states by LU operations in this sense. In addition, we proved that a three-qubit core tensor needs to satisfy only one bi-separability condition to be bi-separable. We further constructed a polytope of three qubits by LU operations that contains all the special states of three qubits that we found.

\section*{Acknowledgement}
The first author is thankful to the numerous discussions with Prof. Isamiddin Rakhimov and Dr. Tay Buang Ann. The first author is sponsored by Universiti Putra Malaysia under the Graduate Research Fellowship (GRF), Ministry of Higher Education Malaysia (MOHE) under MyMaster and Ministry of Education (MOE) under MyBrainSc. This article was supported by the Fundamental Research Grant Scheme (FRGS) under Ministry of Education, Malaysia with project number FRGS/1/2019/STG02/UPM/02/3.

This is a pre-print of an article published in Quantum Information Processing. The final authenticated version is available online at:-
\newline https://doi.org/10.1007/s11128-020-02848-6

\section*{Conflict of interest}
The authors declare that they have no conflict of interest.

\bibliographystyle{ieeetr}
\bibliography{PaperRef}

\end{document}